\documentclass[conference,a4paper]{IEEEtran}
\IEEEoverridecommandlockouts
\usepackage[utf8]{inputenc}
\usepackage[T1]{fontenc}
\usepackage{url}              
\usepackage{cite}             

\usepackage[cmex10]{amsmath}  
\usepackage{amssymb,amsfonts}
\usepackage{dsfont}
\usepackage{mleftright}       
\mleftright                   

\usepackage{graphicx}         
\usepackage{booktabs}         

\usepackage{pgf}

\newcommand{\pow}{\mathsf{P}}

\newcommand{\ind}{\mathbin{\rotatebox[origin=c]{90}{$\vDash$}}}
\newcommand{\hbtoterror}{\textsfP_e}

\newcommand{\textsfR}{{\normalfont \textsf{R}}}
\newcommand{\textsfP}{{\normalfont \textsf{P}}}

\newcommand{\textsfC}{{\normalfont \textsf{C}}}
\newcommand{\textsfV}{{\normalfont \textsf{V}}}
\newcommand{\textsfM}{{\normalfont \textsf{M}}}

\newtheorem{theorem}{Theorem}
\newtheorem{definition}{Definition}
\newtheorem{remark}{Remark}
\renewcommand{\baselinestretch}{0.98}
\begin{document}

\title{Second-order Rate Analysis of a Two-user Gaussian Interference Channel with Heterogeneous Blocklength Constraints}

\author{%
  \IEEEauthorblockN{Kailun Dong, Pin-Hsun Lin, Marcel Mross and Eduard A. Jorswieck}
  \IEEEauthorblockA{%
    Institute for Communications Technology, Technische Universität Braunschweig, Germany\\
    \{k.dong, p.lin, m.mross, e.jorswieck\}@tu-braunschweig.de\\
    }
}

\maketitle

\begin{abstract}
  We consider a two-user Gaussian interference channel with heterogeneous blocklength constraints (HB-GIC), strong interference, and two private messages. We propose to apply the successive interference cancellation with early decoding, i.e., decoding a message with a number of received symbols less than the blocklength at the receiver. We determine the necessary number of received symbols to achieve successful decoding of the longer codeword that satisfies the input power constraints and target average error probability constraints.
  To attain the results, we investigate the dependence testing bound analysis over an independent and identically distributed (i.i.d.) Gaussian input.
  Besides, we derive the second-order achievable rate region of the considered HB-GIC. By numerical results based on the rate-profile approach, we compare the derived second-order rate region to the first-order one, which shows the rate back-off of the considered model due to the impact of finite blocklength.
\end{abstract}

\section{Introduction}\label{sec:introduction}
Ultra-reliable and low-latency communication (URLLC) aims on providing high reliability and low latency for 5G and related technologies, crucial in applications like intelligent transportation, remote healthcare, and industrial automation \cite{01}. Extensive research supports its aim to achieve sub-millisecond wireless communication, enabling real-time decision-making for scenarios like autonomous vehicles and remote surgery.
To achieve URLLC, various technologies like forward error correction coding \cite{03}, adaptive modulation, and interference management are essential.
In practical communication scenarios that involve large-scale and high communication rate demands, such as live streaming of World Cup matches or mobile communication networks during sudden natural disasters, the allocation of network resources among users with limited bandwidth needs to be considered.

The diverse user requirements for latency and service quality have led to research in heterogeneous blocklength-constrained networks. {This idea was first studied as \textit{static broadcasting} in \cite{shulman_static_2000}, where a broadcast channel with heterogeneous decoding deadlines was considered in a first-order asymptotic setting with only a common message. Later, this concept was extended to a network setting \cite{langberg_beyond_2021}, where each node owns some messages and requires other messages with different decoding deadlines. The concept of \textit{joint time-rate region} therein was introduced to characterize the tradeoff between achievable rates and decoding times. In \cite{nikbakht_dirty_2022}, authors consider the case where the transmitter sends two messages at different time points, and the receiver also has different decoding time constraints for the two messages, for a point-to-point setting.} The second-order rate analysis of Gaussian broadcast channel (GBC) with heterogeneous blocklength constraints has been studied in \cite{24, Marcel, Pin_lin}. The combination of superposition coding and early decoding (ED) with successive interference cancellation (SIC) has been applied in this case with i.i.d Gaussian input \cite{06} and composite shell input \cite{Marcel}, respectively. As a fundamental building block in multi-user information theory, the second-order rate region of the Gaussian interference channel (GIC) was investigated in \cite{Le2015}, where it was shown that not only the capacity but also the dispersion is unaffected by the interference for both users. However, the generalization to heterogeneous blocklength constraints is not straightforward, since one user has to perform SIC with an incompletely received interference codeword. This motivates the application of ED in the setting of a GIC with heterogeneous blocklength constraints (HB-GIC). In contrast to \cite{24}, in HB-GIC, messages from two users are transmitted through their own channels, respectively, and also, messages of different transmitters are independent and do not require superposition coding. Besides, interference signals on the two cross-links are both received by each receiver.

The key idea of ED is that when the channel conditions are favorable, users with higher output SNR can successfully decode with fewer received symbols than the designed blocklength. Therefore, early decoding techniques can improve the latency performance of communication systems.
ED has been applied with traditional first-order asymptotic analysis, for example, in finite-state Markov channel \cite{8} and half-duplex cooperative channels \cite{Azarian}. In addition to \cite{8} and \cite{Azarian}, the concept of ED has also been used in several wireless scenarios scenarios, such as cognitive radio (CR) \cite{Jovicic}, binary input channels under a finite blocklength assumption \cite{Sahin} and short message noisy network \cite{HouJie}.
Note that these examples in the above references are only based on the first-order analysis, but not the second-order analysis (in the finite blocklength regime), which motivates our work.

Our main contribution is as follows. We investigate a two-user Gaussian interference channel with heterogeneous blocklength and very strong interference constraints.
We propose an SIC scheme combined with ED and analyze the errors in the first step of SIC at the stronger user 2 by using the dependence testing (DT) bound \cite{5}. We derive the minimum number of received symbols at the receiver required for a successful ED and the corresponding second-order rate region. By numerical results, we show and discuss the impact of channel gains of the cross-link and the blocklength constraints on the rate region in the second order.

This paper is organized as follows. In Section \ref{sec:system-model}, we introduce the system model and some preliminaries. Section \ref{sec:main-results} introduces our main results, and in Section \ref{sec:numerical} we show the numerical simulation. We conclude in Section \ref{sec:conclusion} and sketch the proof of our main result in Appendix \ref{sec:proof}.

\emph{Notation}$\colon$ Upper/lower case normal letters denote random/deterministic variables. Upper-case calligraphic letters denote sets. The notation $a_i^j$ denotes a row vector $[a_i,\,a_{i+1},\,\ldots,\,a_{j}]$, while $a_1^j$ is simplified to $a^j$. We denote the inner product of two vectors $a^j$ and $b^j$ by $\langle a^j,b^j \rangle$. The probability of event $\mathcal{A}$ is denoted by Pr$(\mathcal{A})$. The expectation and variance are denoted by $\mathds{E}[\cdot]$ and $\mbox{Var}[\cdot]$, respectively. We denote the probability density function (PDF) and cumulative distribution function
(CDF) of a random variable $X$ by $f_X$ and $F_X$, respectively. The random variable $X$ following the
distribution with CDF $F$ is denoted by $\,X\sim \,F$. $\mathrm{Unif}(a,b)$ denotes the uniform distribution between $a\in\mathds{R}$ and $b\in\mathds{R}$. We use $X\ind Y$ to denote that $X$ and $Y$ are stochastically independent. The logarithms used in the paper are all with respect to base 2. We define $\textsfC(x)\triangleq\frac{1}{2}\log(1+x)$. Real additive white Gaussian noise (AWGN) with zero mean and variance $\sigma^2$ is denoted by $\mathcal{N}(0,\sigma^2)$. We denote the indicator function by $\mathds{1}$. We denote the inverse $Q$-function by $Q^{-1}(.)$ and the big-$\mathcal{O}$ notation by $\mathcal{O}(.)$.

\section{System Model and Preliminaries}
\label{sec:system-model}
\subsection{System Model}
We consider a two-user HB-GIC with very strong interference, where each user receives only private messages, and there is no common message shared between them. Without loss of generality, we assume that $n_1>n_2$ and the channel gain $a_{21}$ of the cross-link "$X_1 \rightarrow Y_2$" is greater than the channel gain $a_{12}$ of the cross-link "$X_2 \rightarrow Y_1$", i.e., $a_{21}> a_{12}$. Based on the assumption, we define user 1 as the \textit{weaker user} and user 2 as the \textit{stronger user}.
The blocklengths of user 1 and user 2 are denoted by $n_1$ and $n_2$, respectively. In addition, we consider the very strong interference constraint: $a_{21}\geq 1+\textsfP_2$ and $a_{12}\geq 1+ \textsfP_1$\cite[Ch. 21]{Moser_adv}, which is a sufficient condition, where $\textsfP_1$ and $\textsfP_2$ are used in the coming up definition of the power constraints. The received signals at users 1 and 2 at time $j$ can be equivalently expressed respectively as the following \textit{standard form} \cite{Moser_adv}:
\begin{align}
    \begin{cases}
    Y_{1,j}=X_{1,j}+\sqrt{a_{12}}X_{2,j}+Z_{1,j},\\
    Y_{2,j}=X_{2,j}+\sqrt{a_{21}}X_{1,j}+Z_{2,j}, \quad \forall j\in \left \{ 1,\ldots, n_2 \right \},\label{3.13}
    \end{cases}
\end{align}
and
\begin{align}
    \begin{cases}
    Y_{1,j}=X_{1,j}+Z_{1,j},\\
    Y_{2,j}=\sqrt{a_{21}}X_{1,j}+Z_{2,j}, \quad \:\:\forall j\in \left \{ n_{2}+1,\ldots, n_1 \right \},\label{3.14}
    \end{cases}
\end{align}
where $Z_{1,j}$ and $Z_{2,j}$ are two mutually independent additive white Gaussian noises, denoted by
$Z_{1,j} \ind Z_{2,j}$,
and both are independent and identically distributed (i.i.d.), following a standard normal distribution $Z_{1,j}\sim \mathcal{N}(0,1),\,Z_{2,j} \sim \mathcal{N}(0,1)$, for all $j\in \{1,\ldots,n_1\}$.

The transmitters, as well as the receivers, have perfect knowledge of the channel gains $a_{12}$ and $a_{21}$. The goal of the receiver $k$ is to decode message $m_k \in \mathcal{M}_k$ within at most $n_k$ channel uses ($k = 1,2$).
Here the channel inputs $X_{1}^{n_1}\in\mathcal{F}^{n_1}\subseteq\mathbb R^{n_1}$ and $X_{2}^{n_2}\in\mathcal{F}^{n_2}\subseteq\mathbb R^{n_2}$, while $\mathcal{F}^{n_1}$ and $\mathcal{F}^{n_2}$ are sets of feasible codewords satisfying the upcoming power constraints in (\ref{3.16}) and (\ref{3.17}). The considered code is formally defined as follows:

\begin{definition}
    An $(n_1, n_2, \textsfM_1, \textsfM_2, \epsilon, \mathcal{F}^{n_1},\mathcal{F}^{n_2})$-code for an HB-GIC $\textsfP_{Y_1,Y_2|X_1,X_2}$ consists of:
	\begin{itemize}
		\item Two message sets $\mathcal{M}_1 = \{1,2, \dots, \textsfM_1\}$, $\mathcal{M}_2=\{1,2,\dots,\textsfM_2\}$,
		\item Two encoders $\phi_1:\mathcal{M}_1\rightarrow \mathcal{F}^{n_1}$, $\phi_2:\mathcal{M}_2\rightarrow \mathcal{F}^{n_2}$,
		\item Two decoders $\psi_1:\mathbb R^{n_1}\to \mathcal{M}_1$, $y_1^{n_1}\mapsto \hat{m}_1$;  $\psi_2:\mathbb R^{n_2}\to \mathcal{M}_2$, $y_{2}^{n_2} \mapsto \hat{m}_{2}$.
	\end{itemize}
    Assume the message tuple $(m_1,m_2)$ is uniformly selected from $\mathcal{M}_1 \times \mathcal{M}_2$, and the average system error probability satisfies
 {\small
	\begin{align}
		\hbtoterror &:=\! \frac{1}{\textsfM_1\textsfM_2}\!\sum_{m_1=1}^{\textsfM_1}\sum_{m_2=1}^{\textsfM_2}\Pr(\hat{m}_1\!\neq \!m_1\hspace{0.1cm} \text{or}\hspace{0.1cm}\hat{m}_2\!\neq \!m_2\mid \text{($m_1$, $m_2$) is sent})\nonumber\\
        &\leq \epsilon,  \label{eq:total_error_prob}
	\end{align}
 }
\end{definition}
where $\epsilon$ is the target decoding error probability.

We consider the maximal power constraints on the channel inputs, which correspond to the following feasible sets:
\begin{align}
    \mathcal{F}^{n_1}&=\mathcal{F}_{\mathrm{max}}^{n_1}(\textsfP_1):=\{x_1^{n_1}: ||x_1^{n_1}||^2\leq n_1\textsfP_1\},\label{3.16}\\
    \mathcal{F}^{n_2}&=\mathcal{F}_{\mathrm{max}}^{n_2}(\mathsf{\textsfP_2}):=\{x_2^{n_2}: ||x_2^{n_2}||^2\leq n_2\textsfP_2\},
    \label{3.17}
\end{align}
where $\textsfP_1$ and $\textsfP_2$ are constants, and an encoding error is detected when the generated codeword $x_1^{n_1}$ does not belong to $\mathcal{F}^{n_1}$, or $x_2^{n_2}$ does not belong to $\mathcal{F}^{n_2}$.

\begin{figure}[ht]
    \centering
    \includegraphics[width=0.5\textwidth]{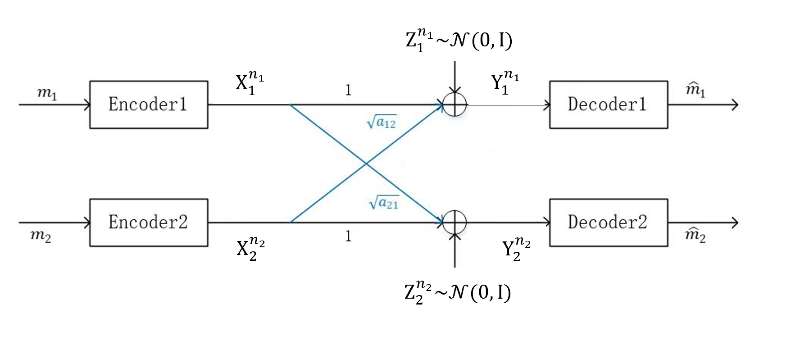}
    \caption{System model of a two-user HB-GIC}
    \label{fig:channel_model}
\end{figure}
We consider that two mutually independent codewords $\{x_1^{n_1}(m_1):m_1\in \mathcal{M}_1\}$ and $\{x_2^{n_2}(m_2):m_2\in \mathcal{M}_2\}$ are both i.i.d. Gaussian generated:
\begin{align}
    \begin{cases}
        X_{1,j}\sim \mathcal{N}(0,\textsfP_1),\qquad \forall j\in \left \{ 1,\cdots, n_1 \right \},\\
        X_{2,j}\sim \mathcal{N}(0,\textsfP_2),\qquad \forall j\in \left \{ 1,\cdots, n_2 \right \}.\nonumber
    \end{cases}
\end{align}

Similar to \cite{Pin_lin}, we consider an SIC decoding scheme based on threshold decoding. At the stronger user, i.e., user 2, the decoder in the first step of SIC finds the smallest \cite{5} $m\in \mathcal{M}_1$, such that $i(x_1^{n_2}(m);y_2^{n_2})>\log \textsfM_1$, where $i(\cdot;\cdot)$ is the information density. If a unique index $m$ is found, set $\hat{m}_1=m$. Otherwise, it declares an error. Based on $\hat{m}_1$, the decoder 2 in the next step finds the smallest $m\in \mathcal{M}_2$, such that $i(x_2^{n_2}(m);\Tilde{y}_2^{n_2})>\log \textsfM_2$, where $\Tilde{y}_2^{n_2}$
is the received signal subtracting the signal $x_1^{n_2}(\hat{m}_1)$, i.e., $\Tilde{y}_2^{n_2}=y_2^{n_2}-x_1^{n_2}(\hat{m}_1)$. If a unique index $m$ is found, set $\hat{m}_2=m$. Otherwise, it declares an error.
At the weaker user, i.e., user 1, the decoder in the first step finds the smallest $m\in \mathcal{M}_2$, such that $i(x_2^{n_1}(m);y_1^{n_1})>\log \textsfM_2$. If a unique index m is found, set $\hat{m}_2=m$. Otherwise, it declares an error. Based on $\hat{m}_2$, the decoder 1 in the next step finds the smallest $m\in \mathcal{M}_1$, such that $i(x_1^{n_1}(m);\Tilde{y}_1^{n_1})>\log \textsfM_1$, where $\Tilde{y}_1^{n_1}$
is the received signal subtracting the signal $x_2^{n_1}(\hat{m}_2)$, i.e., $\Tilde{y}_1^{n_1}=y_1^{n_1}-x_2^{n_1}(\hat{m}_2)$. If a unique index $m$ is found, set $\hat{m}_1=m$. Otherwise, it declares an error.
\subsection{Preliminaries}
We define the successful ED as follows:
\begin{definition}\cite[Def.1]{Pin_lin}
A successful ED means that the stronger user (user 2) with a shorter blocklength constraint can decode messages of the weaker user (user 1), who has a longer blocklength constraint, from the first $\Tilde{n}_1$ received symbols: [$Y_{2,1}, Y_{2,2},\cdots,Y_{2,\Tilde{n}_1}$], where $\Tilde{n}_1 \leq n_2 < n_1$, while the resulting error probability fulfills the target error probability constraint.
\end{definition}
\remark Please note that  without ED, the stronger user (user 2) must wait for $n_1$ received symbols to start the SIC. In contrast, by ED, the SIC can be finished earlier, which reduces the latency.
\remark Please note that user 1 can also decode user 2's message by using ED because of the channel gain $a_{12}\geq 1$, which ensures that the signal achieving user 1's receiver from user 2's transmitter through the cross-link is stronger than its transmission to its own receiver through the direct-link. However, ED does not make sense in this case, though user 1 can decode user 2's message by using symbols less than $n_2$ based on the channel gain assumption, since receiver 1 always needs to wait for all $n_1$ symbols to decode the own wanted message 1. Based on the above reasons, we consider the normal two-step SIC decoding scheme at weaker user 1.

Note that in a P2P Gaussian channel with i.i.d. Gaussian inputs, based on the given channel gain $a$, blocklength $n$, power constraint $\textsfP$, received SNR $a\textsfP$ and target error probability $\epsilon$, there exists the second-order achievable rate \cite{5}, which can be expressed as follows:
\begin{align}
    \textsfR(n,a\textsfP,\epsilon) \leq \textsfC(a\textsfP)-\sqrt{\frac{\textsfV_G(a\textsfP)}{n}}Q^{-1}(\epsilon)+O\left(\frac{\log n}{n}\right),\label{3R}
\end{align}
where the channel dispersion induced by i.i.d. Gaussian input is defined as
$\textsfV_G(a\textsfP) =\log ^2 e\cdot\frac{a\textsfP}{1+a\textsfP}$.
\section{Main Results}
\label{sec:main-results}
In the following, we introduce our main results: under the given target error probability constraint, the minimum number of received symbols required for the successful ED at the stronger user 2 and the second-order rate region.
\vspace{0.5cm}
\begin{theorem}\label{theorem:ed}
Denote $(\epsilon_{1,1}^{\text{SIC}}$, $\epsilon_{1,2}^{\text{SIC}})$ and $(\epsilon_{2,1}^{\text{SIC}}$, $\epsilon_{2,2}^{\text{SIC}})$ as the target decoding error probabilities of $(m_1$, $m_2)$ at the two SIC steps at user 1 and at user 2, respectively, and let $\epsilon$, $ \epsilon_{1}$, and $ \epsilon_{2}$ be the total target decoding error probability and the target decoding error probabilities at user 1 and at user 2, respectively. We assume $n_2 < n_1$, $a_{21}> a_{12}$ and the very strong interference constraint: $a_{21}\geq 1+\textsfP_2$ and $a_{12}\geq 1+ \textsfP_1$. Let the necessary number of symbols to successfully early decode user 1’s signal at user 2 be $\tilde n_1$. If all the following conditions
\begin{IEEEeqnarray}{rCl}
    n_2\geq \tilde n_1 &\geq \frac{\log e \cdot \sqrt{4 \omega_2 \textsfP_1+2 (\omega_2 \textsfP_1)^2} \cdot Q^{-1}\left(\varepsilon_{2,1}^{\text{SIC}}\right)}{2\left(1+\omega_2 \textsfP_1\right) \cdot C\left(\omega_2 \textsfP_1\right)-\log e \cdot \omega_2 \textsfP_1} \cdot \sqrt{n_1}\nonumber\\
    &+\frac{\log \textsfM_1}{C\left(\omega_2 \textsfP_1\right)-\frac{\omega_2 \textsfP_1}{2\left(1+\omega_2 \textsfP_1\right)} \cdot \log e}, \label{TH1}
\end{IEEEeqnarray}
and
\begin{IEEEeqnarray}{rCl}
    \epsilon_{1,1}^{\text{SIC}}+\epsilon_{1,2}^{\text{SIC}}-\epsilon_{1,1}^{\text{SIC}}\epsilon_{1,2}^{\text{SIC}}&&\leq \epsilon_1,\label{483}\\
    \epsilon_{2,1}^{\text{SIC}}+\epsilon_{2,2}^{\text{SIC}}-\epsilon_{2,1}^{\text{SIC}}\epsilon_{2,2}^{\text{SIC}}&&\leq \epsilon_{2},\label{484}\\
    \epsilon_{1}+\epsilon_{2}-\epsilon_{1}\epsilon_{2} &&\leq \epsilon,\label{485}\\
    0< \epsilon_{2,1}^{\text{SIC}},\epsilon_{2,1}^{\text{SIC}},\epsilon_{1,1}^{\text{SIC}}, \epsilon_{1,2}^{\text{SIC}}&&<1,\label{486}
\end{IEEEeqnarray}
are fulfilled, all tuples of rates in $\mathcal{R}_{GIC}$ are achievable, where
\begin{align}
    \mathcal{R}_{GIC}= \Bigg\{ (\textsfR_1, \textsfR_2):
    &\textsfR_1\leq C(\textsfP_1)-\sqrt{\frac{\textsfV_G(\textsfP_1)}{n_1}}Q^{-1}(\epsilon_{1,2}^{\text{SIC}})\nonumber\\
    &\hspace{1.9cm} +O\left (\frac{\log n_1}{n_1} \right),\nonumber\\
    &\textsfR_2\leq C(\textsfP_2)-\sqrt{\frac{\textsfV_G(\textsfP_2)}{n_2}}Q^{-1}(\epsilon_{2,2}^{\text{SIC}})\nonumber\\
    &\hspace{1.9cm} +O\left (\frac{\log n_2}{n_2} \right)\Bigg \},
    \label{Rate-region}
\end{align}
$\textsfV_G:=\log^2 e \cdot \frac{\textsfP_k}{1+\textsfP_k}, k=1,2$
and $\omega_2:= \frac{a_{21}}{1+\textsfP_2}$, $\omega_2 \textsfP_{1}$ is the equivalent output SNR at user $2$ in the first step of SIC.
\end{theorem}
\vspace{0.5cm}
The detailed proof is relegated to the appendix. Here we show the sketch of the proof as follows. Based on the threshold decoding, i.e., the modified typicality decoding, the considered error probability can
be shown to be smaller than a universal bound for all
codewords, which leads to the necessary number of received symbols for a successful ED and the corresponding second-order rate region.

Please note that the conventional SIC used in the first-order analysis of GIC, which requires the whole blocklength
of symbols to decode, may violate the blocklength
constraint. Then, the stronger user may only be able to
use the TIN, which obviously degrades the performance. In contrast, if the cross-link to the stronger receiver is
sufficiently stronger than the direct link to the weaker user, with the aid of
a successful ED, the SIC can still work at the
stronger user while fulfilling the blocklength constraint.

\vspace{0.5cm}
\section{Numerical Results}\label{sec:numerical}
In this section, we first show the performance of latency reduction by using a successful ED. Then we show the second-order achievable rate regions of the SIC with ED schemes as an optimization problem. By using grid search, we finally show the second-order achievable rate region and compare it to the theoretical first-order capacity region.
\subsection{Latency Reduction}
\begin{figure}[htbp]
  \centering
  \includegraphics[width=0.5\textwidth]{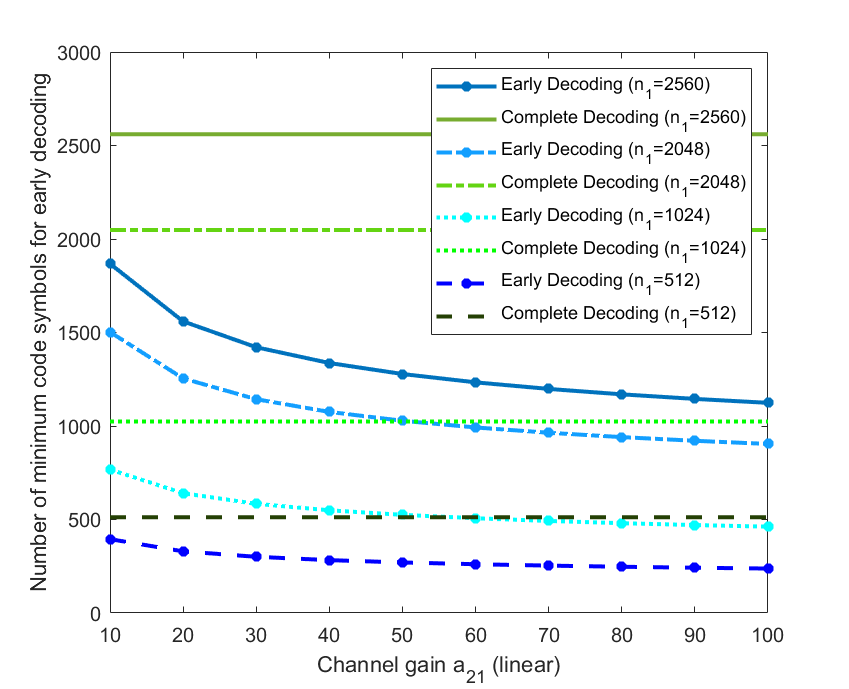}
  \caption{The comparison between two decoding schemes: successful ED with minimum numbers of received symbols from \eqref{TH1} and traditional decoding with complete code symbols.}
  \label{fig:Letency_Reduction}
\end{figure}

We compare the latency between two cases: 1) decoding after the complete codeword is received (without ED); 2) the derived number of received symbols necessary for a successful ED under the second-order rate analysis. We consider the following setting: $\textsfP_1 = 10 $, and $\textsfP_2 = 0.2$. We consider four different blocklengths: $n_1 =512, 1024, 2048,$ and $2560$. They are compared in Fig. \ref{fig:Letency_Reduction}.
Without ED, the stronger user can start to decode only after receiving $n_1$ symbols. When ED is successfully performed, we can observe that the improvement of the latency reduction is enhanced with an increasing $a_{21}$.
\subsection{Second-order rate region}
Due to the possible non-convexity of the rate region of ED, we use the rate-profile method \cite{Mohseni},\cite{Mochaourab}, to find the rate region instead of the common weighted sum rate approach. Then we formulate the problem as follows: given any $\omega,\, 0\leq \omega \leq 1$,
\begin{align}
    \max_{\textsfP_1,\textsfP_2,\epsilon_1,\epsilon_2,\epsilon,\textsfR} & \textsfR \label{ProblemSet}\\
    \mbox{s.t.}\,
    & \textsfR_{1}(\textsfP_1,\epsilon_{1,2}^{\text{SIC}}) \geq \omega \textsfR, \nonumber\\
    & \textsfR_{2}(\textsfP_2,\epsilon_{2,2}^{\text{SIC}}) \geq (1-\omega) \textsfR,\nonumber \\
    & \eqref{TH1},\eqref{483},\eqref{484},\eqref{485},\eqref{486},\eqref{Rate-region}.\nonumber
\end{align}
Note that in the simulation, we invoke the normal approximation by omitting the big-$\mathcal{O}$ terms according to \cite{1}.

\begin{figure}[htbp]
  \centering
  \includegraphics[width=0.5\textwidth]{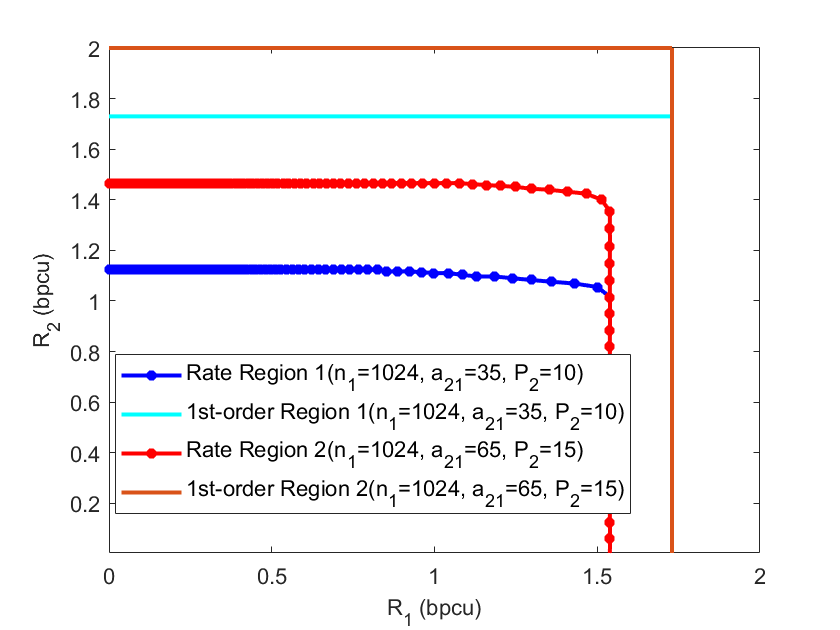}
  \caption{Rate region comparison of ED schemes the HB-GIC. The first-order rate region 1 and second-order rate region 1 use the following setting: $n_1 = 1024, n_2 = 840, a_{12} = 11, a_{21} = 35, \pow_1 = 10, \pow_2 = 10, \epsilon_1 = \epsilon_2 =5\times10^{-5}, \epsilon = 10^{-6}$; the first-order rate region 2 and second-order rate region 2 use the following setting: $n_1 = 1024, n_2 = 840, a_{12} = 11, a_{21} = 65, \pow_1 = 10, \pow_2 = 15, \epsilon = 10^{-6}$.}
  \label{fig:11_35/65_10_10/15_1024}
\end{figure}

\begin{figure}[ht]
  \centering
  \includegraphics[width=0.5\textwidth]{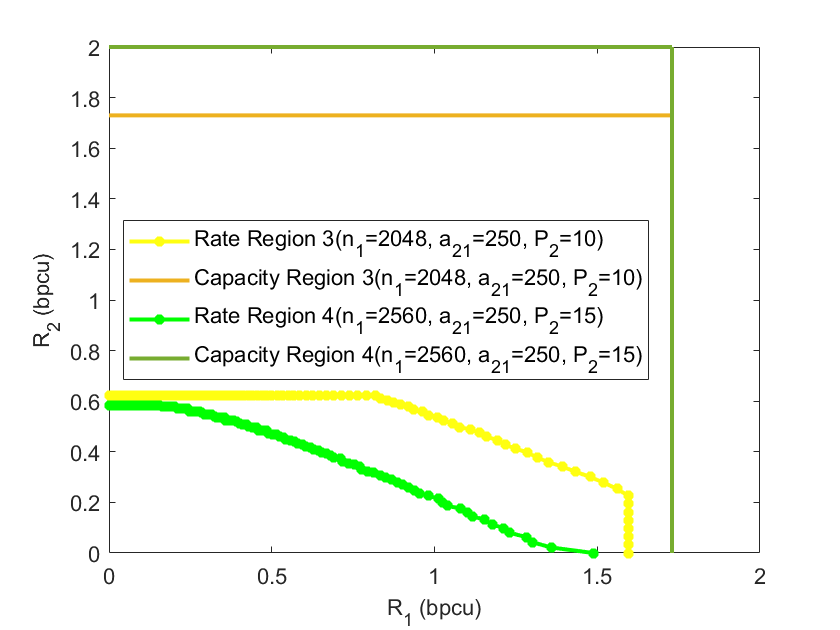}
  \caption{Rate region comparison of ED schemes the HB-GIC. The first-order rate region 3 and second-order rate region 3 use the following setting: $n_1 = 2048, n_2 = 840, a_{12} = 11, a_{21} = 250, \pow_1 = 10, \pow_2 = 10, \epsilon_1 = \epsilon_2 =5\times10^{-5}, \epsilon = 10^{-6}$; the first-order rate region 4 and second-order rate region 4 use the following setting: $n_1 = 2560, n_2 = 840, a_{12} = 11, a_{21} = 250, \pow_1 = 10, \pow_2 = 15,\epsilon = 10^{-6}$.}
  \label{fig:11_250_10_10/15_2048/2560}
\end{figure}

We fix $n_2 = 840$, $a_{12} = 11, \textsfP_1=10,$ and $\epsilon = 10^{-5}$. We consider different blocklengths: $n_1=1024, 2048,$ and $2560$; different power constraints for user 2: $\textsfP_2= 10$ or $15$; also different channel gains: $a_{21}=35,65,$ and $250$. If ED is infeasible, we decrease $\log \textsfM_1$ until ED becomes feasible. The results are shown in Fig. \ref{fig:11_35/65_10_10/15_1024} and \ref{fig:11_250_10_10/15_2048/2560}.
By observing the yellow and green rate region curves in Fig. \ref{fig:11_250_10_10/15_2048/2560}, we notice that, with a fixed channel gain $a_{21}$, in the absence of interference, the second-order rate $\textsfR_1$ for user 1 primarily decreases as the blocklength $n_1$ increases. This is due to the ED constraint \eqref{TH1}. More precisely, when $n_1$ increases, to ensure the rightmost term in \eqref{TH1} is smaller than $n_2$, we need to reduce the feasible region of other variables. This observation is in contrast to the second-order rate region without the ED constraint: if we only consider \eqref{Rate-region} without \eqref{TH1}, increasing $n_1$ and $n_2$ will diminish the impact of the dispersion and leads to a larger rate region. By comparing the red and blue rate regions in Fig. \ref{fig:11_35/65_10_10/15_1024}, while keeping other parameters unchanged, the increase of channel gain $a_{21}$ leads to a noticeable decrease in $\textsfR_2$, while $\textsfR_1$ shows less significant changes. This observation can be attributed to the following reasons: from \eqref{Rate-region}, we can infer that the second-order rate region is primarily influenced by the decoding error probabilities of their respective two-steps SIC processes. Furthermore, because the error probabilities in both the first and second steps of SIC at user 1 and user 2 are subject to their respective target error probability constraints, and to achieve a successful ED at user 2, $n_2$ must satisfy the minimum number of received symbols required for successful ED in \eqref{TH1}. This leads to interdependencies between $n_1$, $n_2$, $a_{21}$, and $\epsilon$, i.e., making changes in blocklength $n_2$ and channel gain $a_{21}$ result in more pronounced variations in $\textsfR_2$ compared to $\textsfR_1$.

\section{Conclusion}
\label{sec:conclusion}
The paper investigated a two-user Gaussian interference channel with heterogeneous blocklengths and very strong interference constraints. We proposed an SIC scheme combined with ED and analyzed the probability of errors by using the DT bound. We characterized the necessary number of received symbols for a successful ED and the achievable second-order rate region. Our numerical results show the interdependency between system parameters and the rate region performance. Our future work will be on improving second-order performance by using a better achievable scheme and also a tight outer bound.
\appendices
\section{Proof of Theorem \ref{theorem:ed}}\label{sec:proof}
To derive the necessary number of received symbols for a successful ED, we investigate the error analysis at the stronger user 2 where $a_{21}>a_{12}$. We consider the following average error probability (at the first step of SIC) at user 2, from the dependence testing bound \cite[Theorem 21]{5} based on a code $\mathcal{C} = \{ x_1^{n_1}(1), x_1^{n_1}(2), \dots, x_1^{n_1}(\textsfM_1)\}$ with blocklength $n_1$, while fulfilling the input power and error probability constraints, where each codeword is Gaussian i.i.d. generated:
\begin{align}
    &\hspace{-0.3cm} \textsfP_e(\mathcal{C}) \leq \frac{1}{\textsfM_1}\!\sum_{m=1}^{\textsfM_1} \bigg \{ \mathds{1}_{x_1^{n_1}(m)\notin\mathcal{F}^{n_1}} \nonumber\\
    &\hspace{0.5cm} + \textsfP_{X_1^{n_2}}\textsfP_{Y_2^{n_2}\mid X_1^{n_2}=x_{1}^{n_2}(m)} \left [  i(x_1^{n_2}(m);Y_2^{n_2})\leq \log \textsfM_1 \right ] \nonumber\\
    &\hspace{0.5cm} +\textsfM_1\textsfP_{X_1^{n_2}}\textsfP_{Y_2^{n_2}}\left [  i(x_1^{n_2}(m);Y_2^{n_2})> \!\log  \textsfM_1\right ] \bigg \},
    \label{4.1}
\end{align}
where we denote $\textsfP_{X_1^{n_2}}\textsfP_{Y_2^{n_2}\mid X_1^{n_2}=x_{1}^{n_2}(m)}$ and $\textsfP_{X_1^{n_2}}\textsfP_{Y_2^{n_2}}$ as the joint probabilities from dependent and independent $(X_1^{n_2}, Y_2^{n_2})$, respectively. In \eqref{4.1}, we consider three error terms: the first term is the probability of cost-constraint violation; the second term is the outage probability given a specific codeword, which means the correct codeword is treated not as transmitted; the third term is the confusion probability given a specific codeword, which means that the wrong codewords are treated as the transmitted ones.

In the first step of SIC at user 2, the received signal can be equivalently expressed as $ \widetilde{Y}_{2,j}:= \sqrt{\omega_2}X_{1,j}+ \widetilde{Z}_{2,j}, \forall j\in \left \{ 1,\cdots, n_2 \right \}$, where $ \widetilde{Z}_{2,j} \sim \mathcal{N} (0,1) $, $ \widetilde{Z}_{2,j} \ind X_{1,j}$, and we define the equivalent channel gain of the first step of SIC at the stronger user as $\omega_2:= \frac{a_{21}}{1+\textsfP_2} $.

Fix any codeword $x_1^{n_2}(m),\, m \in \{ 1, \dots, \textsfM_1\}$ from $\mathcal{C}$, the information density $i(x_1^{n_2}(m); \widetilde{Y}_2^{n_2})$ can be calculated as:
\begin{align}
    i\left(x_1^{n_2}(m) ; \widetilde{Y}_2^{n_2}\right) :=\sum_{j=1}^{n_2} U_j ,
\end{align}
where
\begin{align}
    U_j:=&\textsfC(\omega_2 \textsfP_1)\!+\!\frac{\log e \cdot \omega_2\left(x_{1, j}^2-\textsfP_1 \cdot \widetilde{Z}_{2, j}^2\right)}{2\left(1+\omega_2 \textsfP_1\right)}\nonumber\\
    &\hspace{3cm}+\frac{\log e}{1+\omega_2 \textsfP_1} \sqrt{\omega_2} x_{1, j} \widetilde{Z}_{2, j}.
\end{align}

Then the centralized information density of the $j$-th symbol conditioned on $x_{1,j}(m)$ is as follows:
\begin{align}
    U_j-\mu_j =\frac{\log e}{1+\omega_2 \textsfP_1} \left(\sqrt{\omega_2} x_{1,j} \widetilde{Z}_{2 ,j}+\omega_2 \frac{\textsfP_1}{2}\left(1-\widetilde{Z}_{2, j}^2\right)\right),
\end{align}
where we denote $\mu_j$ as the mean of $U_j$ conditioned on $x_{1,j}$.

We denote the variance of $U_j$ as $\sigma_j^2$, then sum of $U_j$ can be lower bounded as follows:
\begin{align}
    \sum _{j=1}^{n_2}\sigma_j^2 \geq n_2\left( \frac{\log e \cdot \omega_2 \textsfP_1}{\sqrt{2} \cdot \left(1+\omega_2 \textsfP_1\right) }\right)^2.
    \label{4.26}
\end{align}
From \cite{Pin_lin}, the absolute third moment can be upper bounded as follows:
\begin{align}
    &\hspace{0.5cm}\sum\limits_{j=1}^{n_2}\mathbb{E}_{\widetilde{Y}_{2 ,j} \mid X_{1 ,j}=x_{1, j}}
    \left[\left |U_j-\mu_j\right |^3\right]\nonumber\\
    &\hspace{1cm} \leq 4\sum\limits_{j=1}^{n_2}\left(\frac{\log e\sqrt{\omega_2}}{1+\omega_2\textsfP_1}\right)^3\left\{ 2\left |x_{1,j}\right |^3 + 8\left(\sqrt{\omega_2}\textsfP_1\right)^3\right\}\label{4.30}\\
    &\hspace{1cm}:=B_0\left(n_2\right), \label{442}
\end{align}
Note that $B_0(n_2)$ is a constant depending only on $\omega_2, n_2$ and $\textsfP_1$, but not on the realization $x_1^{n_2}$.
The outage probability given $x_1^{n_2}(m)$ can be expressed as follows
\begin{align}
    &\hspace{0.5cm} \textsfP_{ \tilde{Y}_2^{n_2} \mid X_1^{n_2}=x_1^{n_2}}\left[i\left(x_1^{n_2}(m) ; \tilde{Y}_2^{n_2}\right) \leq \log \textsfM_1\right] \nonumber\\
    &\hspace{3cm} \leq Q\left(\gamma _m\left(n_2\right)\right)+6 \cdot B_o\left(n_2\right), \label{451}
\end{align}
where
\begin{align}
    &\hspace{0.5cm}\gamma_m\left(n_2\right)\nonumber\\
    &\geq \frac{2\left(1+\omega_2 \textsfP_1\right) \cdot\left[n_2 C\left(\omega_2 \textsfP_1\right)-\log \textsfM_1\right]-\log e \cdot \omega_2 n_2  \textsfP_1}{\log e \cdot \sqrt{4 \omega_2\left\|x_1^{n_2}(m)\right\|^2+2 n_2 \omega_2^2 \textsfP_1^2}} \label{456}\\
    &:=\gamma _{m, 1}\left(n_2\right).  \label{457}
\end{align}
Besides, according to \cite[Lemma 47]{5}, the confusion probability conditioned on $x_1^{n_2}(m)$ can be upper bounded as follows:
\begin{align}
    \textsfM_1 \cdot \textsfP_{X_1^{n_2}} \textsfP_{\tilde{Y}_2^{n_2}}\left[i\left(x_1^{n_2}(m) ; \widetilde Y _2^{n_2}\right)>\log \textsfM_1\right] \leq B_1\left(n_2\right), \label{confusion}
\end{align}
where we define $B_1\left(n_2\right):=2 \cdot\left(\frac{\ln 2}{\sqrt{\pi {d_1}^2 \textsfP_1^2 \omega_2 n_2}}+12 B_o\left(n_2\right)\right)$ and $d_1:=\frac{\sqrt{\omega _2}}{1+\omega _2 \textsfP_1}$. Note that $B_1\left(n_2\right)$ is a constant depending only on $\textsfP_1,\, n_2$, and $\omega_2$ but not on the realization $x_1^{n_2}$.

Based on the above analysis of outage and confusion probabilities, the error probability at the first step of SIC can be upper bounded by a sum of a $Q$-function and a constant term. In particular, by substituting (\ref{451}) and (\ref{confusion}) into (\ref{4.1}), we can derive the following result
\begin{align}
    & \textsfP_e(\mathcal{C})\leq \lambda  \nonumber\\
    & +Q\!\Bigg(\!\frac{[2\!(\!1\!+\!\omega_2 \!\textsfP_1\!) C\left(\omega_2 \!\textsfP_1\right)\!-\!\omega_2\! \textsfP_1\!\log e ] n_2\!-\!2\!(\!1\!+\!\omega_2\! \textsfP_1\!) \!\log \!\textsfM_1}{\log e \cdot \sqrt{4 \omega_2 \textsfP_1+2 \omega_2^2 \textsfP_1^2} \cdot \sqrt{n_1}}\!\Bigg)\nonumber\\
    & \leq \epsilon_{2,1}^{\text{SIC}}, \label{476}
\end{align}
where
\begin{align}
    \hspace{-0.9cm}\lambda = 6 B_o\left(\!n_2\!\right)\!+\!B_1\left(\!n_2\!\right)\!+\!e^{-\frac{n_1\delta^2}{2}}\!+\!e^{\frac{-n_1^{(2a)}}{2 \textsfP_1}+\ln \left(2 n_1\right)}.
\end{align}
By taking the inverse $Q$-function in (\ref{476}) we know that
\begin{align}
    &Q^{-1}\left(\epsilon_{2,1}^{\text{SIC}}-\lambda\right) \nonumber\\
    &
    \leq \!\frac{[2(\!1\!+\!\omega_2 \textsfP_1\!) C\left(\omega_2 \textsfP_1\right)\!-\!\omega_2 \textsfP_1\log e ] n_2\!-\!2(\!1\!+\!\omega_2 \textsfP_1\!) \!\log \!\textsfM_1}{\log e \cdot \sqrt{4 \omega_2 \textsfP_1+2 \omega_2^2 \textsfP_1^2} \cdot \sqrt{n_1}}, \label{479}
\end{align}
where (\ref{479}) is because of the monotonic decreasing property of the $Q$-function. Furthermore, due to the continuity of the $Q$-function, we can apply Taylor expansion to the inverse Q-function to get:
\begin{align}
    &Q^{-1}\left(\epsilon_{2,1}^{\text{SIC}}\right)+O(\lambda) \nonumber\\
    &\leq \!\frac{[2(\!1\!+\!\omega_2 \textsfP_1\!) C\left(\omega_2 \textsfP_1\right)\!-\!\omega_2 \textsfP_1\log e ] n_2\!-\!2(\!1\!+\!\omega_2 \textsfP_1\!) \!\log \!\textsfM_1}{\log e \cdot \sqrt{4 \omega_2 \textsfP_1+2 \omega_2^2 \textsfP_1^2} \cdot \sqrt{n_1}}.
\end{align}
Therefore, we can find a lower bound for the blocklength $n_2$ under Gaussian approximation as follows
\begin{align}
    &\hspace{-0.5cm}n_2 \geqslant \frac{\log e \cdot \sqrt{4 \omega_2 \textsfP_1+2 \omega_2^2 \textsfP_1^2} \cdot Q^{-1}\left(\epsilon_{2,1}^{\text{SIC}}\right)}{2\left(1+\omega_2 \textsfP_1\right) \cdot C\left(\omega_2 \textsfP_1\right)-\log e \cdot \omega_2 \textsfP_1} \cdot \sqrt{n_1}\nonumber\\
    &\hspace{2cm}+\frac{\log \textsfM_1}{C\left(\omega_2 \textsfP_1\right)-\frac{\omega_2 \textsfP_1}{2\left(1+\omega_2 \textsfP_1\right)} \cdot \log e},
    \label{4.83}
\end{align}
which completes the proof of \eqref{TH1}. By similar steps to \cite{Pin_lin} with proper modifications, we can derive the rate region $\mathcal{R}_{GIC}$ as \eqref{Rate-region}. Due to limited space, we omit the proof steps.


\bibliographystyle{IEEEtran}
\bibliography{references}


\end{document}